# The Correlation among Software Complexity Metrics with Case Study

## Yahya Tashtoush[1], Mohammed Al-Maolegi[2], Bassam Arkok[3]

## Abstract

*People demand for software quality is growing increasingly, thus different scales for the software are growing fast to handle the quality of software. The software complexity metric is one of the measurements that use some of the internal attributes or characteristics of software to know how they effect on the software quality. In this paper, we cover some of more efficient software complexity metrics such as Cyclomatic complexity, line of code and Hallstead complexity metric. This paper presents their impacts on the software quality. It also discusses and analyzes the correlation between them. It finally reveals their relation with the number of errors using a real dataset as a case study.*

## Keywords

## 1. Introduction

The main goal of any system is the quality so costumers and stakeholders are looking for systems with high quality. In information system, it is very important to find the quality factors and improve them. Software quality is the degree to which software possesses a desired combination of attributes such as maintainability, testability, reusability, complexity, reliability, interoperability and etc. (IEEE, 1992). The major way to examine the quality is by continuous and early evaluation of the project progress. The software must have high quality to meet the needs of business and this is an important goal for the developers of software. Due to the appearance of certain factors that affect the quality of software, many practitioners believe that there is a direct relationship between internal attributes such as cost, effort, LOC, speed or memory and external software product attributes such as functionality, quality, complexity, efficiency, reliability or maintainability. For example, a higher number of code lines will lead to greater software complexity and so on.

The complexity of software effects on maintenance activities like software testability, reusability, understandability and modifiability. Software complexity is defined as "the degree to which a system or component has a design or implementation that is difficult to understand and verify" [1]. All the factors that make program difficult to understand are responsible for complexity. So it is necessary to find measurements for software to reduce the impacts of the complexity and guarantee the quality at the same time as much as possible. Because of that, the important challenge is how to maintain the software quality in light of the required functionalities. Various metrics of the complexity may be conducted for the software by the developers such as Hallstead complexity metric, Line of Code or Cyclomatic complexity metric. In the remainder of this paper, section 2 presents a survey about some of metrics of software complexity. Section 3 shows the motivation. Section 4 presents a case study to appear the correlation between the metrics, and their relation to the number of errors. Section 5 includes a conclusion and future work.

## 2. The Software Complexity Metrics

There are many different metrics that are proposed for complexity. We choose some of the metrics that are most important and popular, because of their impact on the project design and code quality. The selected complexity metrics are as the following:

### 2.1. Line of Code (LOC)
In general, LOC is computed by counting lines of a program codes. It is used mostly to evaluate the size of the software but it is not adequate indication to measure the software complexity [2]. It can be computed in many ways as the following:
- LOC: count lines of code.
- SLOC: count source lines of code.
- CLOC: count comment lines of code.

**Manuscript received April 21, 2014.**
 **Yahya Tashtoush**, Jordan University of Science and Technology.
 **Mohammed Al-Maolegi**, Jordan University of Science and Technology.
 **Bassam Arkok**, Jordan University of Science and Technology.

- S&CLOC: count source lines of code with comment lines.
- BLOC: count blanks lines of code.
- PLOC: count physical lines of code.
- LLOC: count logical lines of code.

LOC can be counted in many ways to find the final count. In 1986, Jones determined various methods:

- Count lines of executable only.
- Count lines of executable and data declarations.
- Count lines of executable, comments and data declarations.
- Count lines of executable, comments, data declarations and Job Control Language (JCL).
- Count lines on an input screen as physical lines.
- Count lines by logical determinants as terminated.

In practice, count process is different perspective for the researchers of software engineering. In 1981, Boehm counted LOC as physical lines and encompassed lines of executable, comments and data declarations. In 1986, Conte et al counted LOC as any line in the program except lines of comment and blanks, so LOC includes lines of data declarations, executable and un-executable [3]. LOC can be used as measure for the complexity of software and it is based on the relationship with the defect density and the autonomy between the programming language and the defect density.

LOC is counted rapidly for the programming language and it is understood easily. Moreover, the measurement of LOC is very good but it also has some of weaknesses. Firstly, it doesn't distinguish between the complexities of lines of code. For example, the code "i=1" dose not differ with the code "i= (++x + max (a,b)) / power(c,d)". Actually, the second code is more complex than the first one but LOC metric just counts the number of lines without consideration to anything else. Secondly, LOC ignores the program structure like branches and jumps, so that the code which has more branches and jumps is equal to the code which has the same number of lines and less branches and jumps. Finally, LOC is counted in different methods so its value may differ from one person to another [4].

## 2.2. Halstead Complexity (HC)

Halstead complexity is software metric which classified as a composite complexity metric. It is introduced in 1977 by Maurice Howard Halstead who clarified that the metric has measurable characteristics of program and the relations among them are similar to physical laws [5]. They reveal that the metric should reflect the expression of algorithms or the implementation in varies languages so these metrics are computed based on code statically [6]. Firstly, it extracts four factors:

- n1 = the number of distinct operators in a program
- n2 = the number of distinct operands in a program
- N1 = the total number of occurrences of the operators
- N2 = the total number of occurrences of the operands

Based on these parameters, some formulas can be computed as the following:

- n = n1+ n2.
- N = N1+ N2.
- V = N* $\log_2$ n.
- D= (n1/2)*(N2/n2).
- L = (2*n2) / (n1* N2), that is corresponded to (1/D).
- I = L * V.

Where n is vocabulary of program, N is program length, V is volume, D is difficulty which refers to the difficulty of writing a program or to understand, L is level and I refers to a program's Intelligent content.

The effort measure is the number of distinguishes made during the preparation of a program and it is computed as:

- E=D*V

It is used to compute programming time required (T) as the following:

- T=E/18

Hallstead chooses 18 based on Stroud number. We choose the volume of Hallstead and use it in the case study. It describes the size of execution of program which is nearly the number of bits required to implement the program [7].

Example for (HC):

*If (k < 2)*
*{*
 *If (k > 3)*
  *x = x*k;*
*}*

- Distinct operators: *if ( ) { } > < = *.*
- Distinct operands: *k 2 3 x.*
- n1 = 10
- n2 = 4
- N1 = 13
- N2 = 7

- D= (10/2)*(7/4) =8.75.
- N=13+7=20
- n=10+4=14
- V=20*log2 14 = 76.15,
- E= 8.75*76.15 = 666.31.

From the previous example, we can say that the difficulty of the program is 8.75 and the number of bits required to implement this program is 76.15 bits and its effort to produce the results is 666.31.

Halstead metric is easy to compute and doesn't need to analyze the logic structure of software. It can be used to expect the bug density. In the other side, it has some weaknesses where the complexity is computed depending on the data not the control flow. The operators and the operands of codes and some of branches and jumps are computed with no distinctions. It is surely that the computing of the branches and jumps are more complicated.

## 2.3. Cyclomatic Complexity (CC)

Cyclomatic complexity is one of the most popular metrics. It is used to measure the complexity of a program by measuring the number of linearly independent paths through the source code [8]. Cyclomatic complexity is very simple to compute. It provides a practical way to determine the maximum number of linearly independent paths in a program. It also allows you to evaluate the quality of the program. The high complexity programs contain more errors and detecting them is more difficult. Cyclomatic complexity is computed using the Control Flow Graph (CFG).

CFG illustrates the cycle of the instruction during execution. In other words, CFG describes how to organize or manage the flows throughout the system. To sketch the CFG of a system, go through the following steps. Firstly, number all statements of a program. Each numbered statement considers as a node which refers to command or decision in the program. Secondly, draw an edge between nodes if the result of the execution of the statement is needed to transfer to the next node [9].

When we know how to represent the order, selection (*if...else and switch*) and iteration form (*for, while, do/while and etc.*) for any program, it's easy to draw the CFG.

*An example for (CFG): Insertion sort algorithm:*

```
0   InsertionSort (A, n)
1       for i = 2 to n {
2           key = A[i]
3           j = i – 1
4           while (j > 0) and (A [j] > key) {
5               A [j+1] = A [j]
6               j = j – 1
7           }
8           A [j+1] = key
9       }
```

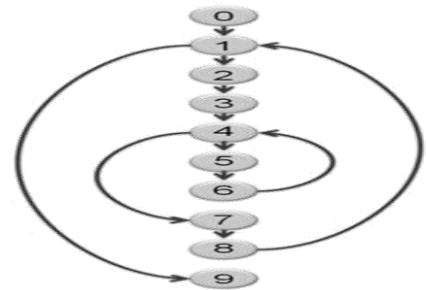

**Figure 1: Control flow graph for insertion sort**

For a control flow graph G of the program, Cyclomatic complexity V (G) can be computed as:

- V (G) = E − N + 2.

Where N is the number of nodes of the control flow graph and E is the number of edges in the control flow graph.

Figure [1] shows that E = 11 and N = 10, therefore Cyclomatic complexity is computed as:

- V (G) = 11 − 10 + 2 = 3.

The program which is measured as a high Cyclomatic complexity will be more difficult to adequately test and more prone to having undiscovered bugs than a simple method with only a few different paths through the code.

In the previous example, the insertion sort algorithm has a Cyclomatic complexity of 3 and this means that there are 3 independent paths through the method. Furthermore it implies that you need at least 3 different test cases to test all the different paths through the code.  If Cyclomatic complexity of program was 15 for example, that is getting difficult to understand, maintain, enhance and reuse.

The experience shows that there is a strong connection between the number of errors and Cyclomatic complexity metric. Cyclomatic complexity should usually be less than 10, and not be more than 20. When Cyclomatic complexity metric is close to 100, the software will be so complicated. Complicated means when an error is fixed, a new error will be revealed with the probability of more than 60%. Thus the code is almost out of control.

Cyclomatic complexity metric is used to measure control flow and ignores the complexity of the data flow of the software. So, measuring a sequential execution codes segment of 10,000 lines using Cyclomatic complexity metric is the same as a single code line. In practice, Hallstead complexity metric and Cyclomatic complexity metric are usually used together. Hallstead complexity metric is used to measure the complexity from the data flow, while Cyclomatic complexity metric measures the control flow.

Cyclomatic complexity metric doesn't differentiate the complexities of different kinds of control flow. In Cyclomatic complexity metric, the complexity of "*if*" is considered as the same with "*case*".

## 3.   Motivation

The behavior of the software product is affected by the internal attributes such as cost, effort or LOC and the external attributes such as functionality, quality or complexity and the relationship between them. The metrics are a combination of these attributes. There are a lot of metrics that are used to manage and control the software product. As the number of used metrics increases, the management and control for software product increases also. But if we ask, is there a relationship between the metrics? If yes, what are the benefits of this relationship? Suppose we have two metrics X and Y. The metrics X is used to measure the complexity so that as the value of X increases, the complexity will increases also and the metrics Y is used to measure the reusability so that as the value of Y increases, the reusability will increases also. The goal is to minimize the value of Y that minimizes the complexity, and maximize the value of X that maximizes the reusability. So, if we study the relationship between the two metrics and find that if the value of X increases, the value of Y increases also. It can be concluded from the relation between the two metrics that the metric X will be used to assess the reusability in addition to the complexity, and the same thing for Y metric that will be used to assess the complexity in addition to the reusability.

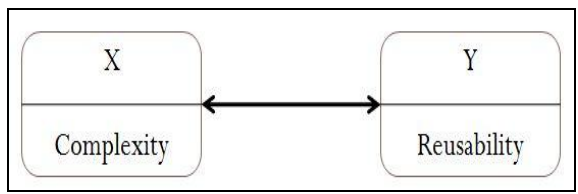

**Figure 2: The relation between the metrics X, Y**

The goal of this paper is to find the correlation between the Cyclomatic metric, Halstead metric and LOC and their correlations with number of errors.

## 4.   Case Study: Finding the Correlation among Complexity Metrics

The target dataset is prepared by selection the relevant features, and then we use the dataset to determine the correlation between Cyclomatic complexity, Hallstead complexity and LOC. After that, we compare between four software projects in term of some of internal and external attributes.

### 4.1. The Dataset
The dataset used in this study are 5 public-domain software defect datasets provided by the NASA IV&V Facility and Metrics Data Program (MDP) repository [10]. The NASA website gives brief descriptions of each MDP dataset:

- CM1: This dataset is from a science instrument written in a C code with approximately 20 kilo-source lines of code (KLOC). It contains 505 modules.
- PC1: This dataset is flight software from an earth orbiting satellite that is no longer operational. It contains 40 KLOC of C code with 1107 modules.
- PC2: This dataset is dynamic simulator for attitude control systems. It contains 26 KLOC of C code with 5589 modules.
- PC3: This dataset is flight software from an earth orbiting satellite that is currently operational. It has 40 KLOC of C code with 1563 modules.
- PC4: This dataset is flight software from an earth orbiting satellite that is currently operational. It has 36 KLOC of C code with 1458 modules.

### 4.2. The Results
The behavior of the software is affected by the internal and external attributes and the relation between them. So, the software attributes must be observed to control the entire software.

We choose arbitrary CM1 dataset that contains 505 modules, and written in C code. We study the correlation between Cyclomatic complexity and Hallstead volume metrics with number of lines of code and number of errors.

From the figures [3, 4], it is obvious that the correlation of Cyclomatic complexity and Hallstead volume with line of code is strong. The change in the number of lines of code will impact on Cyclomatic complexity and Hallstead volume metrics. The correlations as it is obvious in the figures are 0.85 and 0.87 respectively; this indicates a very strong correlation. The regression equation between Cyclomatic complexity and line of code is y = 4.322x - 2.802, the variable x indicates to Cyclomatic complexity, and the variable y indicates to line of code. The regression equation between the Hallstead volume and line of code is y = 0.03x + 1.6724, the variable x indicates to Hallstead volume, and the variable y indicates to line of code.

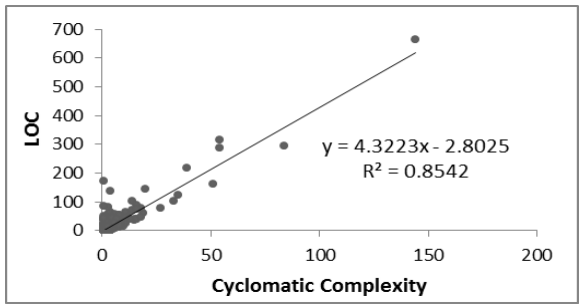

**Figure 3: The correlation between Cyclomatic complexity metric and line of code**

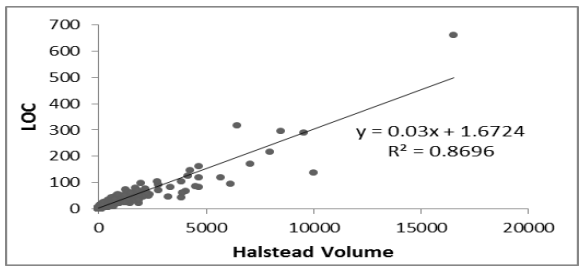

**Figure 4: The correlation between Hallstead volume and line of code.**

From the figures [5, 6], it is obvious that the correlation between Cyclomatic complexity and Hallstead volume with the number of errors is very weak according to this dataset, because each module has a small size and most of them don't contain any errors. The correlations as it is obvious in the figures are 0.059 and 0.033 respectively; this indicates a very weak correlation. The regression equation between Cyclomatic complexity and the number of errors is y = 0.003x - 0.004, the variable x indicates to Cyclomatic complexity, and the variable y indicates to the number of errors. The regression equation between Hallstead volume and the number of errors is y = 3E-05x + 0.0015, the variable x indicates to Hallstead volume, and the variable y indicates the number of errors.

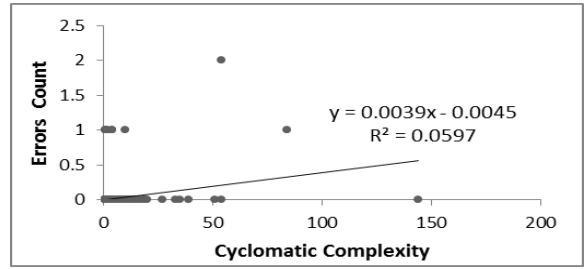

**Figure 5: The correlation between Cyclomatic metric and number of errors**

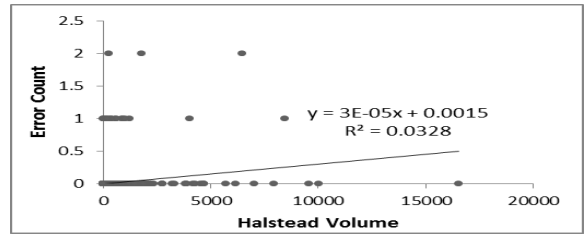

**Figure 6: The correlation between Hallstead volume metric and number of errors**

From the figure [7], it is obvious that that the correlation between Cyclomatic complexity and Hallstead volume is strong. The change in Cyclomatic complexity will impact on Hallstead volume and vice versa. The correlation as it is obvious in the figure is 0.66 and this indicates a strong correlation. The regression equation is y = 114.4x - 98.79, the variable x indicates to Cyclomatic complexity, and the variable y indicates to Hallstead volume.

Cyclomatic complexity and Hallstead complexity are usually used together. Hallstead complexity is used to measure the complexity from the data flow, while Cyclomatic complexity measures the complexity from the control flow.

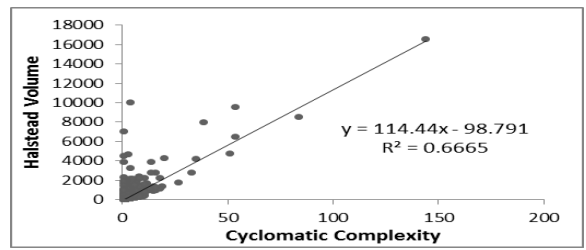

**Figure 7: The correlation between Cyclomatic complexity and Hallstead volume**

In the table [1] below, we put a brief summary of some internal and external attributes of the dataset.

We can determine the software which has the highest or the lowest number modules, line of code, or complexity, and so on.

**Table 1: A brief summary of some internal and external attributes**

|  | Pc1 | Pc2 | Pc3 | Pc4 | CM1 |
|---|---|---|---|---|---|
| # of Modules | 1107 | 5589 | 1563 | 1458 | 505 |
| LOC | 25922 | 26863 | 36473 | 30055 | 16903 |
| Sum of Errors | 139 | 26 | 259 | 367 | 70 |
| Avg of Errors | 0.126 | 0.005 | 0.166 | 0.257 | 0.139 |
| Avg of HV | 700.9 | 104.4 | 786.0 | 543.9 | 862.37 |
| Avg of CC | 5.52 | 1.72 | 5.50 | 4.84 | 5.18 |

## 5.   Conclusion and Future Work

As this paper shows, Cyclomatic complexity has strong correlation with Hallstead complexity and line of code, and weak correlation with the number of errors and this is because the small size of modules of the dataset used in the case study where most of them don't contain any errors. The same thing is for Hallstead complexity which has strong correlation with Cyclomatic complexity and weak correlation with line of code. The correlation between Cyclomatic complexity and Hallstead complexity is strong and they are used together, Cyclomatic complexity to measure the control flow, whereas Hallstead complexity to measure the data flows.
Our future work includes the increasing of dataset, study the relationships between object oriented metrics and other types of metrics, and also study the correlation of web application metrics.

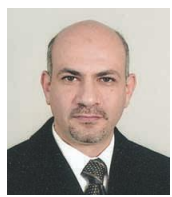

**Dr. Yahya Tashtoush** is an assistant professor in the department of computer science at Jordan University of Science and Technology University in Jordan. He obtained his Ph.D. in Computer Engineering from The University of Alabama In Huntsville in 2006. He received his master in Electrical Engineering from Jordan University of Science and Technology in 1999. He had Bsc. in Electrical Engineering from Jordan University of Science and Technology in 1995. His research interests include: software engineering, wireless networks, artificial intelligence and robotics.

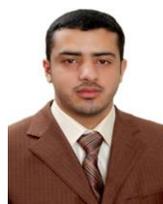

**Mohammed Al-Maolegi** obtained his Master degree in computer science from Jordan University of Science and Technology University in Jordan in 2014. He received his B.Sc. in computer information system from Mutah University in Jordan in 2010. His research interests include: software engineering, software metrics, data mining and wireless sensor networks.

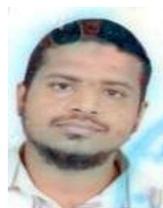

**Bassam Arkok** obtained his Master degree in computer science from Jordan University of Science and Technology University (Jordan) in 2014. He received his B.Sc. in computer science from Alhodidah University (Yemen). His research interests include: software engineering, software metrics, data mining and social networks.